\documentclass[twocolumn,floatfix,showpacs,prd,aps,tightenlines,amsmath,amsfonts,amssymb]{revtex4}

\usepackage{amsmath}
\usepackage{epsfig}
\usepackage{graphicx}
\usepackage{psfrag}
\usepackage{color}
\usepackage{dcolumn}
\usepackage{bm}
\usepackage{longtable}
\usepackage[mathscr]{eucal}
\usepackage{mathrsfs}

\newcommand{\bea}{\begin{eqnarray}}
\newcommand{\eea}{\end{eqnarray}}
\newcommand{\beq}{\begin{equation}}
\newcommand{\eeq}{\end{equation}}

\begin{document}

\def\fun#1#2{\lower3.6pt\vbox{\baselineskip0pt\lineskip.9pt
  \ialign{$\mathsurround=0pt#1\hfil##\hfil$\crcr#2\crcr\sim\crcr}}}
\def\lap{\mathrel{\mathpalette\fun <}}
\def\gap{\mathrel{\mathpalette\fun >}}
\def\kms{{\rm km\ s}^{-1}}
\def\vk{V_{\rm recoil}}


\title{Accuracy of the Post-Newtonian Approximation for Extreme-Mass Ratio Inspirals
from Black-hole Perturbation Approach}

\author{Norichika Sago}
\affiliation{Faculty of Arts and Science, Kyushu University, Fukuoka 819-0395, Japan}
 
\author{Ryuichi Fujita}
\affiliation{CENTRA, Departamento de F\'isica, Instituto Superior T\'ecnico,
Universidade de Lisboa, Avenida Rovisco Pais 1, Portugal}

\author{Hiroyuki Nakano}
\affiliation{Department of Physics, Kyoto University, Kyoto 606-8502, Japan}
\affiliation{Center for Computational Relativity and Gravitation,\\
Rochester Institute of Technology, Rochester, New York 14623, USA}

\begin{abstract}

We revisit the accuracy of the post-Newtonian (PN) approximation
and its region of validity for quasi-circular orbits 
of a point particle in the Kerr spacetime,
by using an analytically known highest post-Newtonian order gravitational
energy flux and accurate numerical results
in the black hole perturbation approach. 
It is found that regions of validity become larger for higher PN order results 
although there are several local maximums in regions of validity 
for relatively low-PN order results. 
This might imply that higher PN order calculations are also encouraged 
for comparable-mass binaries. 

\end{abstract}

\pacs{04.25.Nx, 04.25.dg, 04.30.Db, 95.30.Sf}
\maketitle

\section{Introduction}

To understand binary systems in General Relativity analytically,
we require perturbative techniques. The post-Newtonian (PN) approximation
based on a perturbative expansion by assuming slow motions
and weak gravitational fields of physical systems,  
is one of the most successful approaches
(see e.g., Refs.~\cite{Blanchet:2013haa,Sasaki:2003xr} for a review).
In this approximation, formally $1/c$ where $c$ is the speed of light, 
is considered as a small parameter
that goes to zero in the Newtonian limit $c \to \infty$.
This PN approach has been used to construct accurate
inspiral waveforms of binaries
which are one of the important gravitational wave sources
for the second generation gravitational wave detectors such as
Advanced LIGO (aLIGO)~\cite{TheLIGOScientific:2014jea}, 
Advanced Virgo (AdV)~\cite{TheVirgo:2014hva}, 
KAGRA~\cite{Somiya:2011np,Aso:2013eba}.

The orbital separation of binary systems decreases gradually by the radiation
reaction due to the emission of gravitational waves. As the orbital separation
decreases, the PN approximation will be inappropriate because 
the slow-motion/weak-field assumptions are violated. Therefore, 
we have a basic question: 
``How far can we push the PN approximation in the fast-motion/strong-field?''

In this paper, we treat extreme mass ratio inspirals (EMRIs) as binary systems
to discuss the region of validity of the PN approximation.
For the above systems
which are one of the main targets for a space-based mission,
eLISA Ref.~\cite{Seoane:2013qna},
the black hole perturbation approach is applicable.
In this approach, we consider a point particle with mass $\mu$
is orbiting around a black hole with mass $M$, where $\mu \ll M$.
As an example, in the leading order with respect to the $\mu/M$ expansion,
we consider a quasi-circular orbit of the particle in the Kerr spacetime
with Kerr parameter $a$.
Due to the gravitational radiation reaction, 
the particle will reach the innermost stable circular orbit (ISCO) finally.
In the case of equatorial orbits in Kerr spacetime,
the orbital radius is obtained as~\cite{Bardeen:1972fi}
\begin{align}
 \frac{r_{\rm ISCO}}{M} =& 3 + Z_2 \mp \sqrt{(3 - Z_1) (3 + Z_1 + 2 Z_2)} \,;
 \cr
 Z_1 =& 1 + (1 - (a/M)^2)^{1/3}
 \cr & \times [(1 + a/M)^{1/3} + (1 - a/M)^{1/3} ] \,,
 \cr
 Z_2 =& \sqrt{3 (a/M)^2 + Z_1^2} \,,
\label{eq:ISCO_r}
\end{align}
(we use $c=G=1$ in this paper). 
Here, the upper and lower signs refer to the direct and retrograde orbits
(with $0 \leq a \leq M$), respectively.
We have $r_{\rm ISCO}/M=1$ (direct) and $9$ (retrograde) for $a/M=1$.
The gravitational field along this orbit is so strong
that the PN approximation has the potential to be inappropriate.
The evolution after the ISCO will be described by the ``inspiral to plunge''
transition model developed by \cite{Buonanno:2000ef,Ori:2000zn,Kesden:2011ma}.

There are some works on the region of validity of the PN approximation
in the case of EMRIs. Poisson~\cite{Poisson:1995vs} demonstrated that the
convergence of the PN formula of the energy flux seems to be poor for $v>0.2$.
In the asymptotic analysis of the PN energy flux,
Yunes and Berti~\cite{Yunes:2008tw} found that the edges of the region 
of validity decrease with PN orders beyond 3PN by using the 5.5PN order result,
and the optimal asymptotic approximation is the 3PN order result.
It is possible to consider the PN series as a divergent asymptotic series
(for example, see Ref.~\cite{BenderOrszag} and Section 2 of Ref.~\cite{Yunes:2008tw}). 
This work has been extended to the case of quasi-circular orbits
in the Kerr spacetime by Zhang, Yunes and Berti~\cite{Zhang:2011vha}.

Here, one of the authors of this paper derived the 14PN order energy flux
for a test particle in a circular orbit around a Schwarzschild black
hole~\cite{Fujita:2011zk}. Until this work,
the highest PN order computation was up to the 5.5PN order~\cite{Tanaka:1997dj}.
According to the 14PN result in Fig.~1 of Ref.~\cite{Fujita:2011zk},
we can see a better convergent behavior in the higher PN order
than Fig.~1 of Ref.~\cite{Mino:1997bx}.
Ref.~\cite{Fujita:2011zk} has been extended to even higher PN order,
i.e., 22PN order for circular orbits around a Schwarzschild
black hole~\cite{Fujita:2012cm}
and 11PN for circular orbits around a Kerr black hole~\cite{Fujita:2014eta}.
Recently, the 4PN calculation for general inclined orbits with the sixth order
of the eccentricity has been done in Ref.~\cite{Sago:2015rpa}.
Therefore, it is worth while to revisit 
the region of validity of the PN approximation with this higher PN order result.

The paper is organized as follows. In Sec.~\ref{sec:ERV}, we review
Refs.~\cite{Yunes:2008tw,Zhang:2011vha} and summarize some definitions
used in our paper. The state-of-the-art PN calculation for the Kerr case
in Ref.~\cite{Fujita:2014eta} and the computation of the numerical energy flux
in Refs.~\cite{Fujita:2004rb,Fujita:2009uz} for circular orbit
in the Kerr spacetime are used to obtain an improved estimation. 
Although we use the region of validity defined in Ref.~\cite{Yunes:2008tw} here,
the optimal asymptotic expansion depends on the highest PN order used
in the current analysis. Therefore, in Sec.~\ref{sec:EAR}, we introduce 
another approach based on the error analysis, an eccentricity estimation
to discuss the region of validity. Again, this estimation also depends
on the allowance of error. 
Finally, Sec.~\ref{sec:dis} is devoted to discussions. 
The 22PN results for the Schwarzschild case~\cite{Fujita:2012cm}
is briefly discussed in Appendix~\ref{app:22PN}. 
In this paper, we use the geometric unit system, where $G=c=1$,
with the useful conversion factor 
$1 M_{\odot} = 1.477 \; {\rm{km}} = 4.926 \times 10^{-6} \; {\rm{s}}$.

\section{The edge of region of validity}\label{sec:ERV}

When we discuss the orbital evolution of binaries as a quasi-circular inspiral motion,
we need to know the orbital energy and gravitational energy flux.
In the black hole perturbation approach where we treat the evolution of EMRIs,
the orbital energy is given by the geodesic motion of a particle with mass $\mu$
orbiting around a black hole with mass $M$
in the leading order with respect to the $\mu/M$ expansion.
The energy flux is calculated by the linear-order perturbation 
with respect to $\mu/M$ about the background black hole spacetime.

For the Schwarzshild background, we have a single master equation
for the metric perturbation for so-called odd (axial) and even (polar) 
parity parts, respectively. Regge and Wheeler~\cite{Regge:1957td} 
derived the equation for odd parity perturbation, and later
Zerilli~\cite{Zerilli:1971wd} discussed the even parity part.
On the other hand, for the Kerr background, 
Teukolsky~\cite{Teukolsky:1973ha} derived a differential equation
for perturbations by using the Newman-Penrose formalism~\cite{Newman:1961qr}.
The Teukolsky equation can be solved by the decomposition of $\Psi_4$ as
\begin{align}
 \Psi_4 =& {1\over (r-i a \cos\theta)^{4}}\,
 \displaystyle \sum_{\ell,m}\int_{-\infty}^{\infty}
 \frac{d\omega}{\sqrt{2\pi}} \, 
 e^{-i \omega t + i m \varphi}
 \cr &
 \times {}_{-2}S_{\ell m}^{a\omega}(\theta)\,R_{\ell m\omega}(r) \,,
\end{align}
where $_{-2}S_{\ell m}^{a\omega}$ 
is the spin ($s=-2$) weighted spheroidal harmonic,
and then, the radial function $R_{\ell m\omega}$
has the asymptotic form at the horizon as
\begin{align}
 R_{\ell m \omega}(r\rightarrow r_+) =
 Z^{\rm H}_{\ell m \omega}\,\Delta^2\,e^{-i k r^*} \,,
\end{align}
and at infinity as
\begin{align}
 R_{\ell m \omega}(r\rightarrow \infty)=
 Z^\infty_{\ell m \omega}\,r^3\,e^{i \omega r^*} \,,
\end{align}
where $\Delta=r^2-2Mr+a^2$,
$r_+=M+\sqrt{M^2-a^2}$, $k=\omega-ma/(2Mr_+)$
and $r^*$ denotes the tortoise coordinate
of the Kerr spacetime.
The details to calculate the amplitudes, $Z^{\rm H}_{\ell m \omega}$
and $Z^\infty_{\ell m \omega}$,
are summarized, e.g., in Ref.~\cite{Fujita:2014eta}.

Using the above formulation, the energy flux is derived in two ways,
the analytic and numerical approaches.
In the analytic approach, we consider a series expansion with respect to 
the orbital velocity, $v \ll 1$, i.e., the PN approximation.
When we compute the energy flux numerically, the value is obtained ``exactly''
in the numerical accuracy without any PN approximation.

With the results from the two approaches,
we discuss the region of validity of the PN approximation 
for the gravitational energy flux $F_{g}$
which is related to the loss of orbital energy $E$ as $dE/dt = - F_{g}$,
based on Refs.~\cite{Yunes:2008tw,Zhang:2011vha}.
Here, our notation is the followings. 
In the PN approximation,
the loss of orbital energy normalized by the Newtonian flux $F^{(N)}$
is written in the following form.
\begin{align}
 \frac{dE^{(N)}}{dt} = - F_{\rm Newt} F^{(N)} \,,
 \label{eq:EF}
\end{align}
where $F_{\rm Newt}$ is the Newtonian flux given by
\begin{align}
 F_{\rm Newt} = \frac{32}{5} \left(\frac{\mu}{M}\right)^2 v^{10} \,,
\end{align}
and $F^{(0)}=1$. 
The normalized ``exact'' numerical energy flux 
is denoted by $F$. 
The $N$th-order expansion with respect to $v$
is related to the ($N/2$)PN result.

The analysis of the edge of the region of validity
is originated from Eq.~(19) of Ref.~\cite{Yunes:2008tw},
\begin{align}
 {\cal O}(F-F^{(N)}) = {\cal O}(F^{(N+1)}-F^{(N)}) \,.
\end{align}
This means that the edge is defined by the velocity $v$ at which
the {\em true} error in the PN approximation, $F-F^{(N)}$,  
becomes comparable to the {\em series truncation} error,
$F^{(N+1)}-F^{(N)}$.

\begin{widetext}
\begin{table*}[!ht]
\caption{Approximate edge of the region of validity for
 PN formulas of the energy flux with different PN orders
 in the Schwarzschild ($q=a/M=0$) and
 Kerr ($q=0.1,\,0.3,\,0.5,\,0.9$) cases.
 Here we take the tolerance as $\delta=0.001$.
 For each $N$th-order PN
 approximant, the edge of the region of validity, $\bar{v}^{(N)}$,
 and the relative error evaluated at the edge,
 $\delta F^{(N)}=|F-F^{(N)}|/F$, are shown.
 In this table, we show only the first three figures for each value
 because the more figures are not important for this estimation.}
\label{tab:validity}
\begin{ruledtabular}
\begin{tabular}{|c|c|c|c|c|c|c|c|c|c|c|}
& \multicolumn{2}{c|}{$q=0~~~$} & \multicolumn{2}{c|}{$q=0.1~~~$}
& \multicolumn{2}{c|}{$q=0.3~~~$} & \multicolumn{2}{c|}{$q=0.5~~~$}
& \multicolumn{2}{c|}{$q=0.9~~~$} \\ \cline{1-11}
$N$ & $\bar{v}^{(N)}$ & $\delta F^{(N)}$ & $\bar{v}^{(N)}$ & $\delta F^{(N)}$
& $\bar{v}^{(N)}$ & $\delta F^{(N)}$ & $\bar{v}^{(N)}$ & $\delta F^{(N)}$
& $\bar{v}^{(N)}$ & $\delta F^{(N)}$ \\
\hline
2 & 0.108 & $1.51 \times 10^{-2}$ & 0.108 & $1.47 \times 10^{-2}$
& 0.108 & $1.40 \times 10^{-2}$ & 0.108 & $1.36 \times 10^{-2}$
& 0.111 & $1.33 \times 10^{-2}$ \\
3 & 0.138 & $2.90 \times 10^{-3}$ & 0.137 & $2.83 \times 10^{-3}$
& 0.134 & $2.66 \times 10^{-3}$ & 0.132 & $2.46 \times 10^{-3}$
& 0.129 & $1.98 \times 10^{-3}$ \\
4 & 0.140 & $1.12 \times 10^{-3}$ & 0.141 & $1.19 \times 10^{-3}$
& 0.142 & $1.33 \times 10^{-3}$ & 0.143 & $1.48 \times 10^{-3}$
& 0.145 & $1.82 \times 10^{-3}$ \\
5 & 0.190 & $6.15 \times 10^{-3}$ & 0.191 & $6.09 \times 10^{-3}$
& 0.192 & $6.09 \times 10^{-3}$ & 0.194 & $6.22 \times 10^{-3}$
& 0.201 & $7.04 \times 10^{-3}$ \\
6 & 0.295 & $1.96 \times 10^{-2}$ & 0.313 & $2.95 \times 10^{-2}$
& 0.266 & $9.92 \times 10^{-3}$ & 0.247 & $5.71 \times 10^{-3}$
& 0.233 & $3.07 \times 10^{-3}$ \\
7 & 0.222 & $1.57 \times 10^{-4}$ & 0.223 & $2.49 \times 10^{-4}$
& 0.224 & $4.34 \times 10^{-4}$ & 0.226 & $6.21 \times 10^{-4}$
& 0.229 & $9.98 \times 10^{-4}$ \\
8 & 0.249 & $2.84 \times 10^{-3}$ & 0.249 & $2.78 \times 10^{-3}$
& 0.250 & $2.73 \times 10^{-3}$ & 0.252 & $2.75 \times 10^{-3}$
& 0.258 & $3.01 \times 10^{-3}$ \\
9 & 0.281 & $3.36 \times 10^{-3}$ & 0.285 & $3.85 \times 10^{-3}$
& 0.295 & $5.23 \times 10^{-3}$ & 0.309 & $7.88 \times 10^{-3}$
& 0.356 & $2.72 \times 10^{-2}$ \\
10 & 0.290 & $1.55 \times 10^{-3}$ & 0.291 & $1.38 \times 10^{-3}$
& 0.297 & $1.18 \times 10^{-3}$ & 0.337 & $3.17 \times 10^{-3}$
& 0.290 & $2.99 \times 10^{-4}$ \\
11 & 0.301 & $1.56 \times 10^{-3}$ & 0.300 & $1.49 \times 10^{-3}$
& 0.300 & $1.46 \times 10^{-3}$ & 0.301 & $1.53 \times 10^{-3}$
& 0.304 & $1.92 \times 10^{-3}$ \\
12 & 0.308 & $1.41 \times 10^{-3}$ & 0.311 & $1.58 \times 10^{-3}$
& 0.315 & $1.93 \times 10^{-3}$ & 0.319 & $2.28 \times 10^{-3}$
& 0.328 & $2.99 \times 10^{-3}$ \\
13 & 0.377 & $1.59 \times 10^{-2}$ & 0.371 & $1.19 \times 10^{-2}$
& 0.371 & $9.93 \times 10^{-3}$ & 0.376 & $1.06 \times 10^{-2}$
& 0.399 & $1.77 \times 10^{-2}$ \\
14 & 0.422 & $3.01 \times 10^{-3}$ & 0.375 & $1.15 \times 10^{-3}$
& 0.388 & $2.16 \times 10^{-3}$ & 0.419 & $6.66 \times 10^{-3}$
& 0.354 & $2.01 \times 10^{-5}$ \\
15 & 0.341 & $3.19 \times 10^{-4}$ & 0.343 & $4.66 \times 10^{-4}$
& 0.347 & $7.56 \times 10^{-4}$ & 0.351 & $1.04 \times 10^{-3}$
& 0.362 & $1.73 \times 10^{-3}$ \\
16 & 0.361 & $2.52 \times 10^{-3}$ & 0.361 & $2.30 \times 10^{-3}$
& 0.363 & $2.15 \times 10^{-3}$ & 0.367 & $2.26 \times 10^{-3}$
& 0.393 & $4.63 \times 10^{-3}$ \\
17 & 0.367 & $1.28 \times 10^{-3}$ & 0.371 & $1.61 \times 10^{-3}$
& 0.381 & $2.46 \times 10^{-3}$ & 0.396 & $4.18 \times 10^{-3}$
& 0.394 & $1.27 \times 10^{-3}$ \\
18 & 0.387 & $2.60 \times 10^{-3}$ & 0.390 & $2.50 \times 10^{-3}$
& 0.392 & $1.78 \times 10^{-3}$ & 0.403 & $1.56 \times 10^{-3}$
& 0.381 & $7.69 \times 10^{-4}$ \\
19 & 0.390 & $1.07 \times 10^{-3}$ & 0.386 & $8.21 \times 10^{-4}$
& 0.384 & $7.04 \times 10^{-4}$ & 0.383 & $7.99 \times 10^{-4}$
& 0.381 & $1.21 \times 10^{-3}$ \\
20 & 0.378 & $5.29 \times 10^{-4}$ & 0.380 & $6.93 \times 10^{-4}$
& 0.383 & $1.01 \times 10^{-3}$ & 0.387 & $1.33 \times 10^{-3}$
& 0.391 & $1.96 \times 10^{-3}$ \\
21 & 0.396 & $2.55 \times 10^{-3}$ & 0.396 & $2.31 \times 10^{-3}$
& 0.397 & $2.19 \times 10^{-3}$ & 0.401 & $2.40 \times 10^{-3}$
& 0.411 & $3.67 \times 10^{-3}$ \\
\end{tabular}
\end{ruledtabular}
\end{table*}

\end{widetext}

\begin{figure}[!ht]
 \includegraphics[width=0.48\textwidth,clip=true]{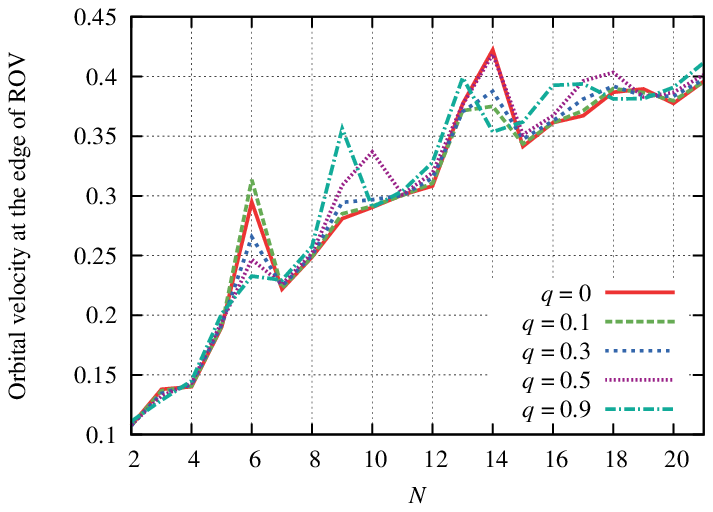}
 \includegraphics[width=0.48\textwidth,clip=true]{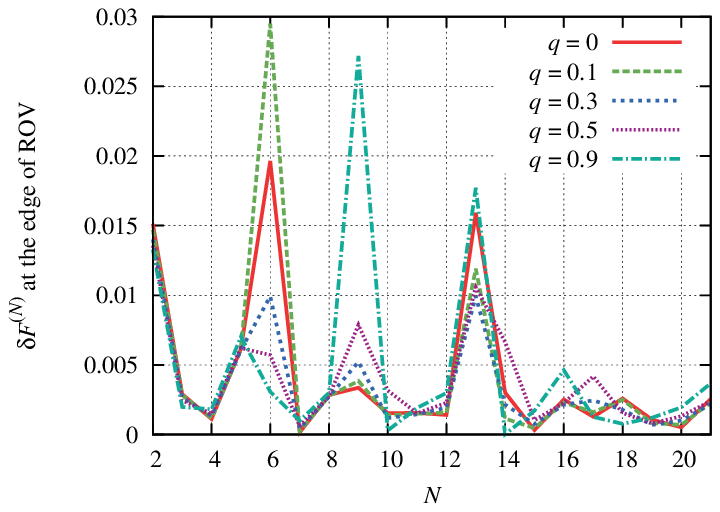}
\caption{
  {\it Top}: Approximate values of the region of validity (ROV) in the velocity, 
  $\bar{v}^{(N)}$ vs. $N$ for $\delta=0.001$
  in the Schwarzschild ($q=a/M=0$) and Kerr ($q=0.1,\,0.3,\,0.5,\,0.9$) cases.
  {\it Bottom}: The relative error evaluated at the edge,
  $\delta F^{(N)}=|F-F^{(N)}|/F$.}
\label{fig:N_vs_ROV}
\end{figure}

In order to evaluate the above equation,
it would be the easiest way to introduce a tolerance.
According to Eq.~(20) of Ref.~\cite{Yunes:2008tw},
\begin{align}
 \Delta^{(N)}_{{\rm ROV}} \equiv
 \left||F-F^{(N)}| - |F^{(N+1)}-F^{(N)}|\right| < \delta \,,
 \label{eq:ERV}
\end{align}
where $\delta$ is some tolerance,
we define the approximate edge of the region of validity.

Since the estimation of the edge depends on the tolerance,
an uncertainty width obtained from variations of the tolerance
has been shown in Ref.~\cite{Yunes:2008tw}.
For simplicity, we focus only on a constant tolerance.
In Table~\ref{tab:validity} and Fig.~\ref{fig:N_vs_ROV},
we present ``approximate'' values of 
the edge of the region of validity evaluated from Eq.~(\ref{eq:ERV})
with $\delta=0.001$ which is the same value used in their paper,
for the Schwarzschild ($q=a/M=0$)
and Kerr ($q=0.1,\,0.3,\,0.5,\,0.9$) cases.
Here, we show the approximate edge of the region of validity
in terms of the orbital velocity,
\begin{align}
 v =& (M\Omega)^{1/3} \,; \cr
 \Omega =& \frac{M^{1/2}}{r^{3/2}+aM^{1/2}} \,,
 \label{eq:v_r}
\end{align}
and the relative error,
\begin{align}
\delta F^{(N)} = \frac{|F-F^{(N)}|}{F} \,,
\end{align}
evaluated at the edge.
Up to $N=10$ for the Schwarschild case ($q=a/M=0$),
this is consistent with Table I 
of the erratum of Ref.~\cite{Yunes:2008tw}.

It is noted that in the Schwarzschild case
the 3PN ($N=6$) order gives the optimal approximation
in the analysis up to the 5.5PN order calculation ($N=10$ here).
We find a similar feature even in the Kerr cases with small spins
($q=0.1$ and $0.3$)
if we consider only the $N=8$ calculations.
Although there are various local peaks in Fig.~\ref{fig:N_vs_ROV},
we have larger regions of validity basically
if the higher PN results are introduced.
In the Kerr case with the spin $q=0.9$,
we obtain a large region of validity for $N=9$.

We also note the velocity at the ISCO~\cite{Bardeen:1972fi}
which is calculated by substituting Eq.~(\ref{eq:ISCO_r})
in Eq.~(\ref{eq:v_r}) as $r=r_{\rm ISCO}$.
In practice, the ISCO velocity is derived as
$v \approx 0.408248$ ($q=0$),
$0.418954$ (0.1),
$0.444210$ (0.3),
$0.477084$ (0.5),
and $0.608618$ (0.9).
When we just treat the ISCO velocity as a reference,
the edge of the region of validity
presented in Fig.~\ref{fig:N_vs_ROV} does not reach
the ISCO velocity for all cases.
However, for example, Ref.~\cite{Fujita:2011zk} showed that the 14PN
gravitational waveforms
can extract an accurate physical information from two-years observation of EMRIs
in the Laser Interferometer Space Antenna (LISA) band.
Therefore, we need to take account of physical (observational)
situations to discuss the practical validity of approximations.

In Ref.~\cite{Yunes:2008tw}, an appropriate tolerance estimation,
$\delta_N(v) = |F^{(N+2)}(v)-F^{(N+1)}(v)|$
which depends on the PN order,
has been introduced because higher-order approximations should
be evaluated by an appropriate smaller tolerance.
The edge of the region of validity is different from the approximate one
obtained in the above.
Then, they searched the edge by
\begin{align}
\Delta^{(N)}_{{\rm ROV}} < \delta_N(v) \,.
\label{eq:ERV2}
\end{align}
In the first step, the velocity is set at $v=0.2$, i.e.,
the tolerance is $|F^{(N+2)}(0.2)-F^{(N+1)}(0.2)|$.
By using $\bar{v}^{(N)}$ evaluated by this tolerance,
the appropriate tolerance in the next iteration is determined as
$|F^{(N+2)}(\bar{v}^{(N)})-F^{(N+1)}(\bar{v}^{(N)})|$.
Although we obtain a consistent result with Table II of the erratum
of Ref.~\cite{Yunes:2008tw} where they used three successive
iterations, we find that there is no appropriate
convergent solution for the edge in many iteration.
In practice, when we consider $v=0.2$ as the initial guess for the iteration,
we have the solution of $\bar{v}^{(N)} \to 0$, or
$\bar{v}^{(N)} > v_{\rm ISCO}$
which is beyond our analysis.
These behaviors depend on the existence of an intersection
of two curves, $\Delta^{(N)}_{{\rm ROV}}$
and $\delta_N$ as functions of $v$, and the solutions mean that
the inequality of Eq.~(\ref{eq:ERV2}) holds everywhere
in the case of $\bar{v}^{(N)} > v_{\rm ISCO}$ for the quasi-circular evolution.
Therefore, we do not extend this analysis for higher PN orders here.

\section{Edges of the allowable region}\label{sec:EAR}

One of the simplest analysis for the region of validity
is to give an allowance $\delta$ for the difference
between the numerical (exact) and analytic (PN) results as
\begin{align}
|F-F^{(N)}| < \delta \,.
\end{align}
In practice, $\delta$ will be numerical accuracies, 
for example, the allowance of the constraint violations
in numerical relativity (NR), and so on.
In NR simulations of binary black hole
systems~\cite{Pretorius:2005gq, Campanelli:2005dd, Baker:2005vv},
the gravitational waveforms 
are one of the most important output.
There are various requirements to obtain the accurate waveforms. 
One of them is to keep minimizing constraint violations.
Since we have discussed the instantaneous valid region
at each velocity, this can be considered
as the constraint violations of each time slice in the NR simulations
\footnote{Strictly speaking, the PN formula used in this paper should
not be applied to comparable mass binaries treated in NR because it
does not contain the corrections of the finite mass ratio.
We used it to estimate the PN convergence for comparable mass cases
assuming the convergence does not depend on the mass-ratio strongly.
Ref.~\cite{LeTiec:2011bk} has given a supporting evidence for this
assumption although the relativistic periastron advance, not the energy
flux, is discussed there. Hence we hope that the estimate may be
suggestive of the convergence of the unknown PN terms.}.

The error in the PN approximation can be also expressed as the
deviation from the ``exact'' quasi-circular evolution, {\it i.e.},
the orbital eccentricity, defined by
\begin{align}
 e \equiv& \left(\frac{dE}{dr}\right)^{-1}
 \frac{1}{v}\, F_{\rm Newt}\, |F-F^{(N)}|
 \,,
 \label{eq:ee}
\end{align}
where the orbital energy is given by
\begin{align}
 \frac{E}{\mu} =&
 {\frac {{r}^{3/2}-2\,M\sqrt {r}+a\sqrt {M}}{{r}^{3/4} \left( {r}^{3/2}
 -3\,M\sqrt {r}+2\,a\sqrt {M} \right)^{1/2} }} \,,
\end{align}
and we may convert the orbital radius to the velocity as
\begin{align}
 r = {\frac { \left[  \left( M-{v}^{3}a \right) \sqrt {M} \right]^{2/3}}
 {{v}^{2}}} \,.
\end{align}

To derive Eq.~(\ref{eq:ee}),
we convert the energy flux to the radial orbital evolution
in quasi-circular inspirals,
\begin{align}
 \frac{dr}{dt} =& \left( \frac{dE}{dr} \right)^{-1} \,\frac{dE}{dt} 
 \cr
 =& - \left( \frac{dE}{dr} \right)^{-1} \, F_{\rm Newt}\, F \,.
\end{align}
In the PN approximation, we may use $F^{(N)}$ from Eq.~(\ref{eq:EF}),
instead of the exact $F$ in the above expression.
By analogy to the Newtonian eccentric orbit,
the radial trajectory of the orbit with a small eccentricity is
written as
\begin{align}
 r(t) \sim r_{0}\left[ 1-e \cos \left(\frac{v t}{r_{0}}\right) \right] \,.
 \label{eq:Newtonian-radial}
\end{align}
Here, it is not necessary to treat the difference
between the azimuthal and radial frequencies in the Newtonian approximation
\footnote{To check if the "Newtonian eccentricity" defined in Eq.~(\ref{eq:ee})
works well, we recalculated the edge of the allowable region by replacing the
Newtonian frequency, $v/r_0$ in Eq.~(\ref{eq:Newtonian-radial}) to
the 1PN radial frequency. We got the almost same result as in the
Newtonian case (the difference is less than 1\%). Therefore, we
think the Newtonian eccentricity is enough for the current purpose.}.
In the quasi-circular orbital evolution, 
the evolution of the radius $r_{0}$ is described by using the exact $F$,
and the difference between $F$ and $F^{(N)}$ is related to the eccentricity as
\begin{align}
 \left|\left(\frac{dr}{dt}\right)^{(N)}-\frac{dr}{dt}\right|
 =& \left( \frac{dE}{dr} \right)^{-1} \, F_{\rm Newt}\,|F-F^{(N)}|
\cr 
 =& e v \sin \left(\frac{v t}{r_{0}}\right) \,,
\end{align}
where we do not consider the oscillation, but treat only the amplitude,
i.e., $e v$ in the last equality.
Since $v$ in Eq.~(\ref{eq:ee}) is a overall factor
and we want to just give a rough error estimator,
it is sufficient to calculate the eccentricity as the estimator
by using the circular orbital velocity $v$ in the azimuthal direction.

Here, as a reference, we simply pick up a number from Table 1
in Ref.~\cite{Ajith:2012tt}.
The lowest one is $e \sim 2 \times 10^{-5}$ in SpEC~\cite{SXS_GWD} 
(numerical relativity) waveforms by using an iterative procedure
to reduce the eccentricity~\cite{Boyle:2007ft,Buonanno:2010yk}
(see also Refs.~\cite{Pfeiffer:2007yz,Purrer:2012wy}).
It should be noted that there are various factors
to produce the eccentricity in the initial data.

\begin{figure}[!ht]
 \includegraphics[width=0.48\textwidth,clip=true]{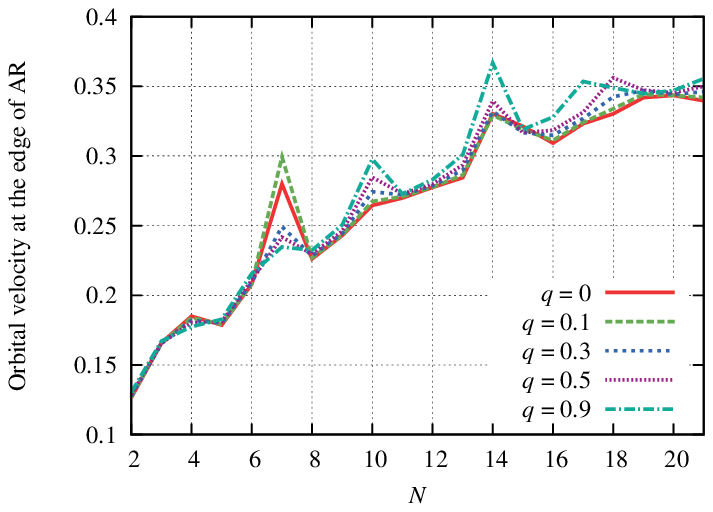}

 \includegraphics[width=0.48\textwidth,clip=true]{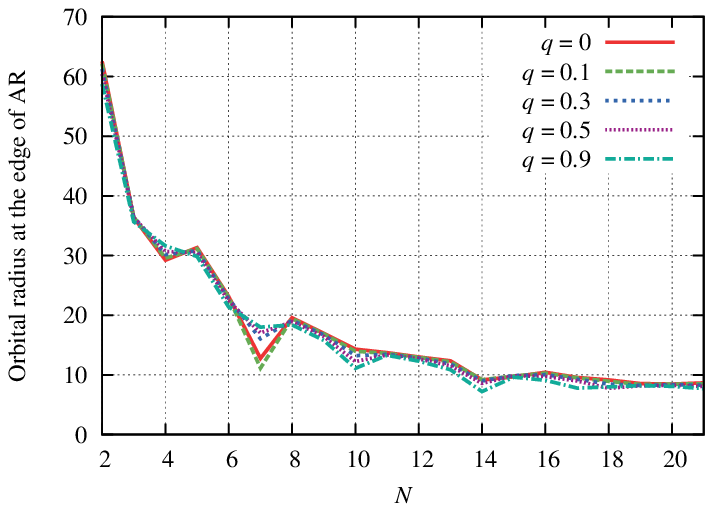}
\caption{
  The edge of the allowable region (AR) from the eccentricity estimation
  in the case of $e \leq 1 \times 10^{-5}$
  in terms of the orbital velocity (top) and radius (bottom).}
\label{fig:N_vs_AR}
\end{figure}

Therefore, for simplicity, we set a restriction on $e$ from the error
in the energy flux as
\begin{align}
 e \leq 1 \times 10^{-5} \,.
 \label{eq:eerest}
\end{align}
Using this restriction, it is found that the edge of the allowable region
is obtained as Fig.~\ref{fig:N_vs_AR}
in terms of the orbital velocity (top), $v^{(N)}$, and radius (bottom)
for the Schwarzschild ($q=0$)
and Kerr ($q=0.1,\,0.3,\,0.5,\,0.9$) cases.

In the case of $v<v^{(N)}$,
we expect that the PN errors do not influence
the eccentricity in the quasi-circular orbital evolution
in the current NR simulations.
We note that an large edge $v^{(N)}$ is seen for $N=7$,
and $N \geq 13$ is required to transcend this value
in the Schwarzschild ($q=0$) case.
Again, we have larger allowable regions basically
if the higher PN results are introduced,
although there are various local peaks in Fig.~\ref{fig:N_vs_AR}.
In the bottom panel of Fig.~\ref{fig:N_vs_AR},
we see that there is a convergent behavior in the orbital radius
at $r \sim 8.5M$
in the restriction of $e = 1 \times 10^{-5}$.

In addition, comparing the top of Fig.~\ref{fig:N_vs_AR} to
the top of Fig.~\ref{fig:N_vs_ROV}, one can find that the global
behaviors of $\bar{v}^{(N)}$ and $v^{(N)}$
are almost same although each value of the velocity or
the locations of the local peaks are slightly different.
This indicates the global property of the PN convergence.

\section{Discussions}\label{sec:dis}

There are various estimates for the region of validity of the PN
approximation.
For example, the NINJA
(Numerical INJection Analysis)-2 project~\cite{Ajith:2012tt} has
established that certain 3PN order gravitational waveforms
are sufficiently accurate for $v \lesssim 0.33$ to use them as templates
for searching gravitational waves from comparable-mass black-hole binaries
with the second generation gravitational
wave detectors such as
Advanced LIGO (aLIGO)~\cite{TheLIGOScientific:2014jea}, 
Advanced Virgo (AdV)~\cite{TheVirgo:2014hva}, 
KAGRA~\cite{Somiya:2011np,Aso:2013eba},
In the NRAR project~\cite{Hinder:2013oqa}, 
the eccentricity $e \lesssim 0.002$ is one of the requirements
for quasi-circular  black-hole binaries.

On the other hand, EMRIs treated in this paper
are one of the targets for space-based gravitational wave detectors.
Although in previous analyses, e.g., Ref.~\cite{Fujita:2011zk}
discussed the 14PN gravitational waveforms
in the Laser Interferometer Space Antenna (LISA) band,
the analysis will be sufficient
for a new plan, eLISA Ref.~\cite{Seoane:2013qna}.

In this paper, using the best-known analytic PN and accurate numerical results,
11PN calculation for circular orbits around a Kerr black hole~\cite{Fujita:2014eta},
we revisited the analysis discussed by Yunes and Berti~\cite{Yunes:2008tw},
and obtained the same results in the 5.5PN order calculation
for the Schwarschild case.
The main results are presented in Table~\ref{tab:validity}
and Fig.~\ref{fig:N_vs_ROV}.
In order for the edge of the region of validity
to reach the ISCO velocity ($v \approx 0.608618$ for $q=0.9$),
much higher PN order calculations
will be required. But, this may be just for mathematical interest,
and we should discuss gravitational waveforms
in practical gravitational wave observations.

Next, we have introduced an eccentricity estimation in Eq.~(\ref{eq:ee})
to discuss the allowable region for each PN order, and 
found that for the small spin cases ($q=0$ and $0.1$)
it was difficult to see the improvement due to the use
of higher PN order calculations in the previous analysis.
This is because the 3.5PN ($N=7$ in Fig.~\ref{fig:N_vs_AR}) order gives
a large allowable region in the $q=0$ and $0.1$ cases
and the higher PN calculation
than 6.5PN ($N \geq 13$ in Fig.~\ref{fig:N_vs_AR})
order is required to extend this region.
On the other hand, 
the advantage at $N=7$ is lost in higher spin cases.
In terms of the orbital radius, 
a comparatively smooth expansion of the allowable region can be seen
in the higher PN order approximation.
Although we have discussed the PN approximation in the case of EMRIs,
the higher PN order calculation for comparable-mass binaries
would be encouraged.

\begin{figure}[!ht]
 \includegraphics[width=0.48\textwidth,clip=true]{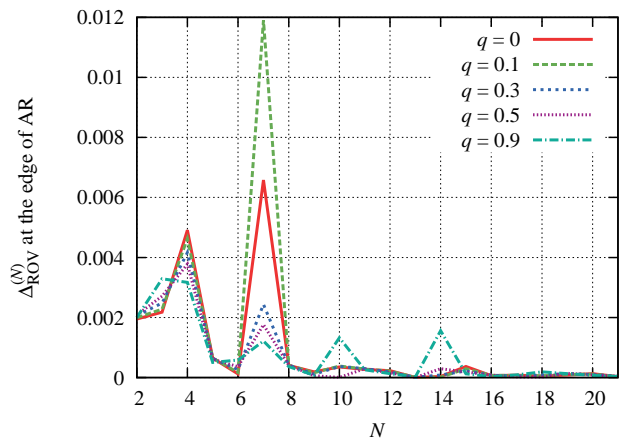}
\caption{
  The value of $\Delta^{(N)}_{{\rm ROV}}$ at the edge of the allowable
  region by using the result shown in Fig.~\ref{fig:N_vs_AR}.}
\label{fig:delta_ROV}
\end{figure}

Here, we note that a reference in Eq.~(\ref{eq:eerest}) has been
assumed to determine the edge of the allowable region for the PN approximation.
This means that the edge largely depends on what to discuss. 
Therefore, it is reasonable
to consider the combination of two analyses presented in this paper,
i.e., the region of validity and the allowable region.
In Fig.~\ref{fig:delta_ROV},
we show $\Delta^{(N)}_{{\rm ROV}}$ (defined in Eq.~(\ref{eq:ERV}))
evaluated at the edge of the allowable region
which is obtained in Fig.~\ref{fig:N_vs_AR}.
It is found that only some higher PN order results
in the analysis of the edge of the allowable region
are valid in the sense of the region of validity
if we introduce a constant tolerance.
For example, for the higher spinning case ($q=0.9$),
we observe that the large allowable region for $N=10$ and $14$ from
the eccentricity estimation does not satisfy the constant tolerance $\delta=0.001$
used for the analysis of the region of validity in Sec.~\ref{sec:ERV}.
Therefore, we conclude that any local peak in the top panels of Figs.~\ref{fig:N_vs_ROV}
and~\ref{fig:N_vs_AR} does not indicate the best approximation
in a given PN order.

In this paper, we focus on the PN convergence of the energy flux,
which corresponds to the dissipative piece of the self-force under
the adiabatic approximation. It is worth considering the PN convergence
of the post-adiabatic effects of the self-force, the conservative
and the second order dissipative self-forces.
In Ref.~\cite{Isoyama:2012bx}, for example, the impact of the
post-adiabatic self-force on
the gravitational waves has been discussed assuming that the convergence
is similar to that of the energy flux. This assumption and their results
should be justified when the post-adiabatic self-force is directly
calculated.

Finally, we have now various numerical and analytical results
for the gravitational radiation reaction
in the black hole perturbation approach.
For example, Ref.~\cite{Fujita:2009us} discussed
a particle moving on eccentric inclined orbits numerically
(see also analytical solutions of the bound timelike geodesic orbits
in the background Kerr spacetime~\cite{Fujita:2009bp}).
The analytical results have been used 
to calculate the factorized waveforms~\cite{Damour:2007xr,Damour:2008gu}
which are employed in the effective-one-body approach~\cite{Buonanno:1998gg}
(see Refs.~\cite{Fujita:2010xj,Pan:2010hz}). 
Studying region of validity in the PN approximation for these cases 
is left for future work. 


\acknowledgments 

We would like to thank Soichiro Isoyama to encourage us to complete
this research. 
N.~S. acknowledges support by JSPS Grant-in-Aid for 
Young Scientists (B), No.~25800154. 
R.~F.'s work was funded through H2020 ERC Consolidator Grant 
"Matter and strong-field gravity: New frontiers in Einstein's theory" 
grant agreement no. MaGRaTh-646597. 
H.~N.'s research was supported by MEXT Grant-in-Aid for Scientific Research
on Innovative Areas,
``New Developments in Astrophysics Through Multi-Messenger Observations
of Gravitational Wave Sources'', No.~24103006. 

\appendix

\section{From 22PN result for Schwarzschild}\label{app:22PN}

From Eq.~(5)
of Ref.~\cite{Fujita:2011zk} in the PN approximation,
the energy flux normalized by the Newtonian flux is written as
\begin{align}
 F^{(N)} = \sum_{k=0}^{N} \sum_{p=0}^{[k/6]} F^{(k,p)} (\ln(v))^p v^k \,,
 \label{eq:formalFN}
\end{align}
where $F^{(0)}=1$, and the normalized ``exact'' numerical energy flux 
is denoted by $F$. 

\begin{figure}[!ht]
\includegraphics[width=0.98\linewidth]{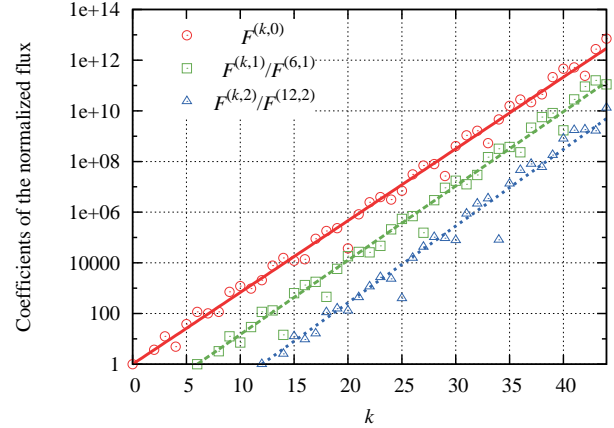}
\caption{The log-linear plot for the absolute magnitude of some coefficients
 in Eq.~(\ref{eq:formalFN}) as a function of PN order.}
\label{fig:higherPNc}
\end{figure}

In this appendix, we use the 22PN order calculation for circular orbits
around a Schwarzschild black hole~\cite{Fujita:2012cm, Fujita:2014eta}.
In Fig.~\ref{fig:higherPNc}, we present $|F^{(k,p)}|$ in Eq.~(\ref{eq:formalFN})
for the cases of $p=0$, $1$ and $2$.
Interestingly, the coefficients are approximately fitted by 
a linear function with respect to $k$ in the log-linear plot.
Then, the fitting functions are obtained as
\bea
F^{(k,0)} &=& (1.91983)^k \,,
\nonumber \\ 
\frac{F^{(k,1)}}{F^{(6,1)}} &=& (1.96395)^{k-6} \,,
\nonumber \\
\frac{F^{(k,2)}}{F^{(12,2)}} &=& (2.00873)^{k-12} \,.
\eea
This fitting means that the radius of convergence
for the orbital velocity is derived as
\bea
v_R^{(p=0)}&\approx&0.520879 \,,
\nonumber \\
v_R^{(p=1)}&\approx&0.509177 \,,
\nonumber \\
v_R^{(p=2)}&\approx&0.497828 \,, 
\eea
from the Cauchy ratio test, respectively.
These values cover the entire physical domain
up to the ISCO velocity, $v_{\rm ISCO} \sim 0.408$.

\end{document}